\documentstyle[11pt]{article} 
\input epsf

\begin{document}

\title{
Phase diagram of polymer blends\\
in confined geometry\footnote{Workshop on Modern Problems in Soft Matter Theory, Lviv, Aug 27-31, 2000,\\
                              submitted to the Journal of Molecular Liquids and Condensed Matter Physics}
}

\author{
M.\ M\"{u}ller$^{1*}$, K.\ Binder$^{1}$, and E.V. Albano$^{2}$
\\
{\small ${}^1$ Institut f{\"u}r Physik, WA 331, Johannes Gutenberg Universit{\"a}t}
\\
{\small D-55099 Mainz, Germany}
\\
{\small ${}^2$ INIFTA, Universidad de La Plata, C.C.\ 16 Suc.\ 4.}
\\
{\small 1900 La Plata, Argentina}
}

\date{\today, draft}
\maketitle

\begin{abstract}
Within self--consistent field theory we study the phase behavior of a symmetrical binary $AB$ polymer blend confined into a thin
film. The film surfaces interact with the monomers via short range potentials. One surface attracts
the $A$ component and the corresponding semi--infinite system exhibits a first order wetting transition.
The surface interaction of the opposite surface is varied as to study the crossover from capillary
condensation for symmetric surface fields to the interface localization/delocalization transition for
antisymmetric surface fields. In the former case the phase diagram has a single critical point close to
the bulk critical point.  In the latter case the phase diagram exhibits two critical points which correspond
to the prewetting critical points of the semi--infinite system. Only below a triple point there is a single two
phase coexistence region.  The cross\-over between these qualitatively different limiting behaviors occurs gradually, 
however, the critical temperature and the critical composition exhibit a non-monotonic dependence on the surface field.

The dependence of the phase behavior for antisymmetric boundaries is studied as a function of the film thickness
and the strength of the surface interactions. Upon reducing the film thickness or decreasing the strength of the
surface interactions we can change the order of the interface localization/delocalization transition from first to
second. 

The role of fluctuations is explored via Monte Carlo simulations of a coarse grained lattice model.
Close to the (prewetting) critical points we observe 2D Ising critical behavior. At lower temperatures capillary waves
of the $AB$ interface lead to a pronounced dependence of the effective interface potential on the lateral system size.

{${}^*$ email: \tt{Marcus.Mueller@uni-mainz.de}}
\end{abstract}

\section{Introduction}
Confining a binary mixture gives rise to a rich interplay between wetting and miscibility behavior.\cite{EREV,PREV,DIETRICH,NAKANISHI}
In a porous material or a slit--like pore with identical surfaces the critical point of the mixture is shifted away from its bulk value
to lower temperatures and higher compositions of the component prefered by the walls.\cite{NAKANISHI} The phenomenon of capillary condensation
which occurs if the two confining boundaries are symmetric is well--known.\cite{GUBBINS} In some applications (e.g., coatings or dielectrics), however, the surfaces 
of the film (e.g., a solid substrate and vacuum) interact very differently with the constituents of the mixture. In the opposite limit of antisymmetric
walls, i.e., one wall attracts the $A$--component with exactly the opposite strength than the other wall the $B$--component, an interface 
localization/delocalization transition\cite{BROCHARD,PE,SWIFT,BINDER,NEW} occurs close to the wetting transition of the semi--infinite system.

We study the dependence of the phase behavior of a polymer mixture confined into a thin film on the surface interactions and film thickness via 
self--consistent field calculations\cite{MSCF1,MSCF2,MSCF3} and Monte Carlo 
simulations.\cite{MCSCF} Our paper is arranged as follows: In the next section we describe our mean field calculations and discuss the dependence
of the phase diagram on the monomer--wall interactions. Then we investigate the effect of fluctuations via simulations of a coarse 
grained lattice model. The paper closes with a discussion of the anticipated chain length dependence and an outlook on experimental systems.

\section{Self--consistent field calculations}
We calculate the phase behavior of a confined $AB$ mixture within the self-consistent field theory of  Gaussian 
polymers.\cite{SCF,MARK} The film comprises a volume $V_0=\Delta_0 \times L \times L$. $\Delta_0$ denotes 
the film thickness, while $L$ is the lateral extension of the film. The density at the film surfaces decreases 
to zero in a boundary region of width $\Delta_w$ according to\cite{FILM} 
\begin{equation}
\Phi_0(x) = \left\{ \begin{array}{ll}
               \frac{1-\cos\left( \frac{\pi x}{\Delta_w}\right)}{2} 
	&	\mbox{{\small ; $0 \leq x \leq \Delta_w$}} \\
	       1                                                   
		& \mbox{{\small ;  $\Delta_w \leq x \leq \Delta_0 - \Delta_w$}} \\
               \frac{1-\cos\left( \frac{\pi (\Delta_0 - x)}{\Delta_w}\right)}{2}  
		 & \mbox{{\small ; $\Delta_0 - \Delta_w \leq x \leq \Delta_0$}}
             \end{array}\right. 
	     \label{eqn:dens}
\end{equation}
where $\Phi_0$ denotes the ratio of the monomer density and the value $\rho$ in the middle of the film.
The thickness $\Delta$ of an equivalent film with constant monomer density $\rho$ is 
$\Delta=\Delta_0-\Delta_w$. We choose $\Delta_w=0.15 R_e$,\cite{FILM} where $R_e$ is the end--to--end distance.
Both surfaces interact with the monomer species via a short range potential $H$:
\begin{equation}
H(x) = \left\{ \begin{array}{ll}
            \frac{4 \Lambda_1 R_e\left\{1+\cos\left( \frac{\pi x}{\Delta_w}\right)\right\}}{\Delta_w} 
	    & \mbox{{\small ; $0\leq x \leq \Delta_w$ }} \\
            0                                                                                         
	    & \mbox{{\small ;  $\Delta_w \leq x \leq \Delta_0 - \Delta_w$}} \\
            \frac{4 \Lambda_2 R_e\left\{1+\cos\left( \frac{\pi (\Delta_0 - x)}{\Delta_w}\right)\right\}}{\Delta_w}
	    & \mbox{{\small ; $\Delta_0 - \Delta_w \leq x \leq \Delta_0$}}
        \end{array}\right.
\end{equation}
$H>0$ is attractive for the $A$ monomers and repulsive for the $B$ species.
The normalization of the surface fields $\Lambda_1$ and $\Lambda_2$, which act on the monomers close to the 
left and the right surface, is chosen such that the integrated interaction energy between the surface and the 
monomers is independent of the width of the boundary region $\Delta_w$.

$A$ and $B$ polymers contain $N$ monomers and are structurally symmetric.
The polymer conformations $\{ {\bf r}_\alpha(\tau)\}$ determine the microscopic $A$ monomer density 
$\hat \Phi_A({\bf r}) = \frac{N}{\rho} \sum_{\alpha=0}^{n_A} \int_0^1 {\rm d}\tau \delta\left({\bf r}-{\bf r}_\alpha(\tau)\right)$,
where the sum runs over all $n_A$ $A$ polymers in the system and $0 \leq \tau \leq 1$ parameterizes the 
contour of the Gaussian polymer. A similar expression holds for $\hat \Phi_B({\bf r})$.
With this definition the semi--grandcanonical partition function takes the form:
\begin{eqnarray}
{\cal Z} &\sim & \sum_{n_A=1}^n \frac{e^{+\Delta \mu n_A/2k_BT}}{n_A!} \;\;
                              \frac{e^{-\Delta \mu n_B/2k_BT}}{n_B!} 
               \int {\cal P}_A[{\bf r}] 
               \int {\cal P}_B[{\bf r}]  \;\delta\left( \Phi_0 - \hat \Phi_A - \hat \Phi_B \right) \nonumber \\
         &&       \;\times \;\exp\left( - \rho \int {\rm d}^3{\bf r} 
                \left\{ \chi \hat \Phi_A \hat \Phi_B - H(\hat \Phi_A-\hat \Phi_B) \right\}\right)   
\end{eqnarray}
where $n=n_A+n_B$ and $\Delta \mu$ represents the exchange potential between $A$ and $B$ polymers.
The functional integral ${\cal P}$ sums over all conformations of the polymers with the statistical weight
$\exp \left(- \frac{3}{2R_e^2} \int_0^1 {\rm d}\tau \;\left(\frac{{\rm d}{\bf r}}{{\rm d}\tau}\right)^2 \right)$
of a non--interacting Gaussian polymer.  The second factor 
enforces the monomer density profile to comply with Eq.(\ref{eqn:dens}) (incompressibility). 
The Boltzmann factor in the partition function incorporates the 
thermal repulsion between unlike monomers, which is described by the Flory--Huggins parameter $\chi$,  and the interactions between  monomers and surfaces. 

In mean field approximation the free energy is obtained as the extremum of the semi--grandcanonical free energy functional:
\begin{eqnarray}
\frac{{\cal G}}{n k_BT}  \equiv  +\; \ln\frac{n}{V_0}  - \;\ln {\cal Q}           
     + \; \frac{1}{V} \int {\rm d}^3{\bf r} \;\; \chi N \Phi_A \Phi_B                      
                                              -  H N \left\{ \Phi_A-\Phi_B\right\}\nonumber \\
     - \; \frac{1}{V} \int {\rm d}^3{\bf r} \;\;  \left\{ W_A \Phi_A + W_B \Phi_B\right\} +  \Xi \left\{\Phi_0 -  \Phi_A -  \Phi_B\right\} 
						 \label{eqn:F}
\end{eqnarray}
with respect to its arguments $W_A,W_B,\Phi_A,\Phi_B,\Xi$. ${\cal Q}_A$ denotes the single chain partition function:
\begin{equation}
{\cal Q}_A[W_A] = \frac{1}{V_0} \int {\cal D}_1[{\bf r}] {\cal P}_1[{\bf r}] \;\; e^{ - \int_0^1 {\rm d}\tau\; W_A({\bf r}(\tau)) }
\end{equation}
and similarly for ${\cal Q}_B$;
${\cal Q} =  \exp(\Delta \mu/2k_BT) {\cal Q}_A + \exp(-\Delta \mu/2k_BT) {\cal Q}_B$.

The values of $W_A,W_B,\Phi_A,\Phi_B,\Xi$ which extremize the free energy functional are denoted by lower--case letters 
and satisfy the self-consistent set of equations
\begin{equation}
w_A({\bf r}) = \chi N \phi_B({\bf r}) - H({\bf r}) N + \xi({\bf r}) \qquad \mbox{and} \qquad
\phi_A({\bf r}) = -\frac{V}{{\cal Q}} \frac{{\cal DQ}_A}{{\cal D}w_A({\bf r})} 
\label{eqn:scf}
\end{equation}
$\Phi_0({\bf r}) = \phi_A({\bf r}) + \phi_B({\bf r})$, and similar expressions for $w_B$ and $\phi_B$.
Substituting the extremal values of the densities and fields into the free energy functional (\ref{eqn:F}) we 
calculate the free energy $G$ of the different phases. At coexistence the two phases have equal semi--grandcanonical 
potential.

To calculate the monomer density we employ the end segment distribution which satisfies a diffusion equation.
We expand the spatial dependence of the densities and fields in a set of orthonormal 
functions.\cite{MARK,FILM} This procedure results in a set of 
non--linear equations which are solved by a Newton--Raphson--like method. We use up to 
80 basis functions and achieve a relative accuracy $10^{-4}$ in the free energy.

The phase diagram in a thin film ($\Delta_0 = 0.9 R_e$) with antisymmetric surface fields of various strengths
is presented in Fig.1. For weak surface fields we find a second--order interface localization/delocalization transition.\cite{MSCF3}

Close to the critical point of the bulk enrichment layers form gradually at each surface and stabilize an interface in the
center of the film. The films remains laterally homogeneous in this ``soft--mode'' phase. Only below the second order transition 
the interface is localized at one wall and $A$--rich and $B$--rich domains coexist laterally. Upon increasing the monomer--wall 
interaction strength the transition temperature shifts to lower values.

\vspace*{5mm}
\noindent
    \begin{minipage}[t]{126mm}%
       \mbox{
       ({\bf a})
       \hspace*{-1cm}
        \setlength{\epsfxsize}{6.08cm}
        \epsffile{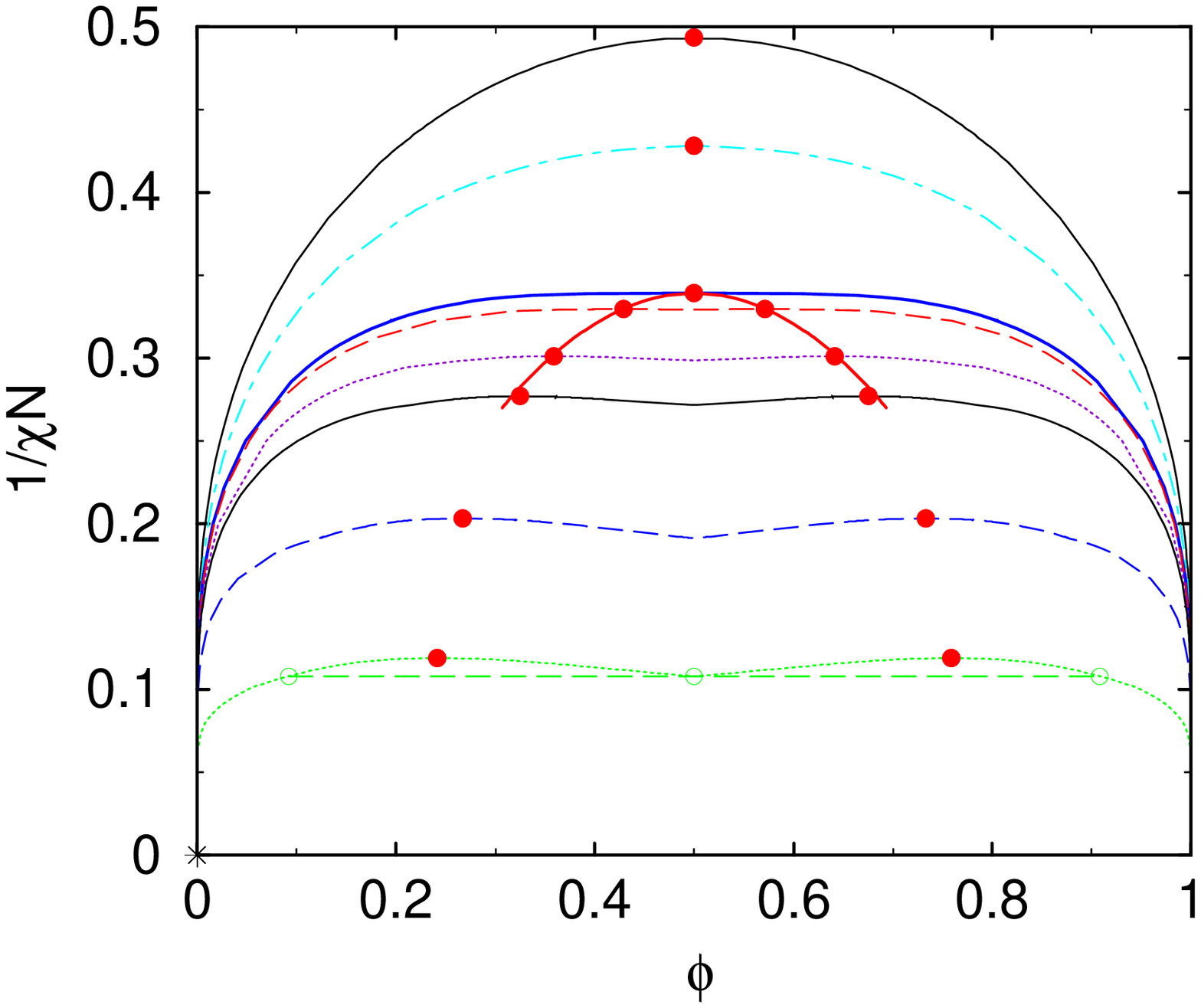}
       ({\bf b})
       \hspace*{-0.7cm}
        \setlength{\epsfxsize}{6.3cm}
        \epsffile{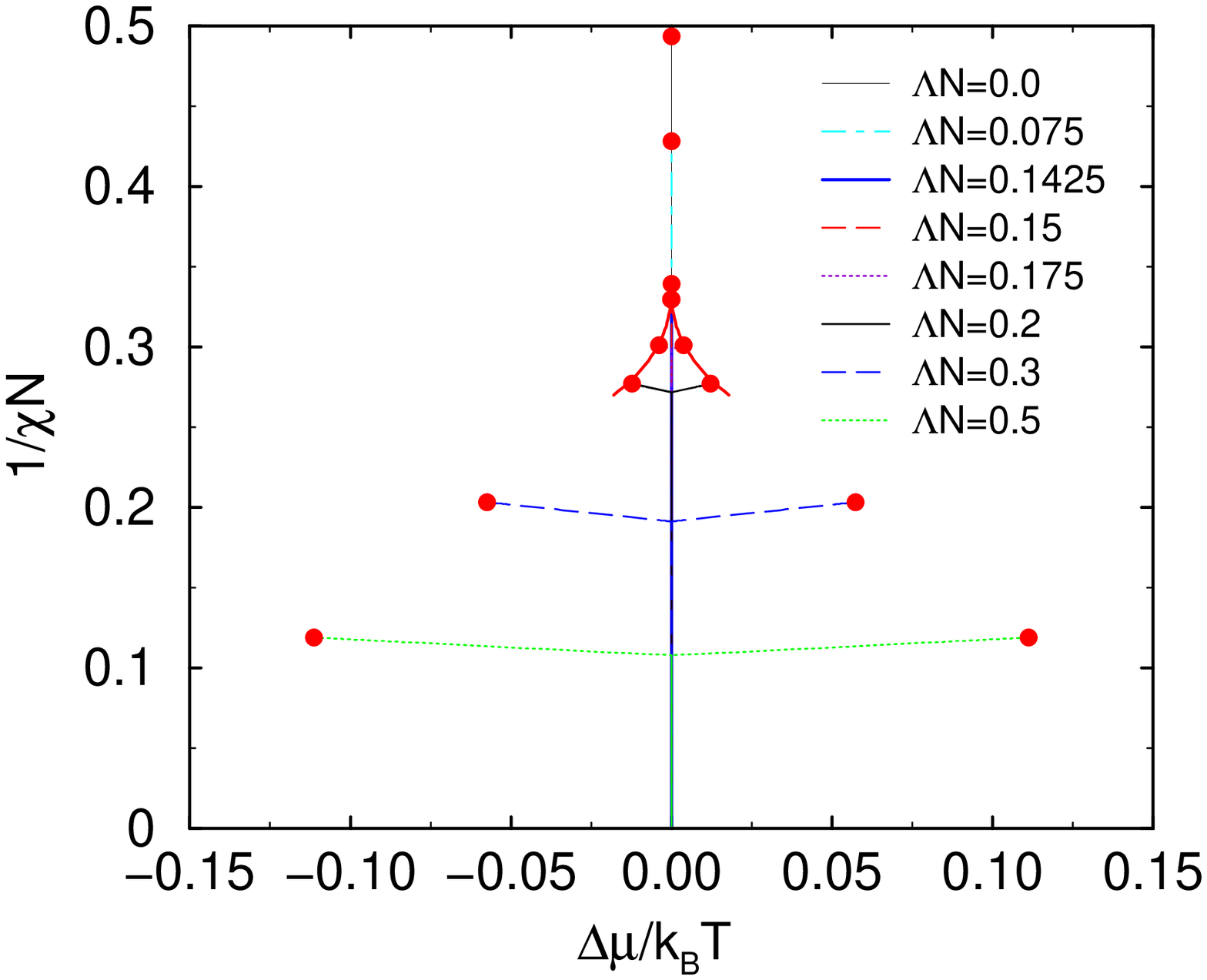}
       }
    \end{minipage}%
    \hfill%
    \begin{minipage}[b]{126mm}%
    {\small 
    Fig.1: ({\bf a}) Phase diagrams in a film with antisymmetric surface fields.
    The values of the surface fields $\Lambda N$ are indicated in the key. For
    $\Lambda N\leq 0.1425$ we find a single critical point, while we find two critical points
    for larger surface fields.
    ({\bf a}) displays the phase diagram in the temperature--composition plane, while ({\bf b})
    presents the coexistence curves $\Delta \mu_{\rm coex}(\chi N)$. (For $\Lambda N=0.15$
    the two critical points are indistinguishable on the scale of panel ({\bf b})). From \protect\cite{MSCF3}.
    }
    \end{minipage}%
\vspace*{5mm}

\noindent

At large monomer--wall interaction the wetting transition is of first order and we observe also a first order interface 
localization/delocalization transition.\cite{MSCF2,MSCF3} The prewetting coexistence slightly above the first order wetting transition temperature
gives rise to two miscibility gaps in a film. At each wall a thin and a thick enrichment layer of the prefered component coexist.
These coexistence regions terminate in two critical points which are the analogs of the prewetting critical points at each wall.
When we decrease the temperature the miscibility gaps open and form a triple point at which an $A$--rich phase, a phase where the 
interface is delocalized in the middle of the film, and a $B$--rich phase coexist. Below the triple temperature an $A$--rich and 
a $B$--rich phase coexist.

The two types of phase diagrams are separated by a tricritical interface localization/delocalization transition. For our model 
we find a tricritical transition at $\Lambda N=0.1425$ and $\Delta_0 = 0.9 R_e$. 
When we increase the film width the tricritical strength of the surface interactions decreases.
At tricriticality, there is only a single critical point, but the binodals are characterized by an exponent $\beta=1/4$ in mean field approximation.

Fig.1b presents the coexistence curves as a function of temperature and exchange potential. For a second order or a tricritical transition
phase coexistence occurs at $\Delta \mu=0$ due to the symmetry of the system with respect to exchanging $A \rightleftharpoons B$. For a first order transition
the coexistence curve lays on the symmetry axis $\Delta \mu=0$ for  low temperatures. When the temperature rises, it bifurcates at the triple point; 
each continuation is the thin film analog of the prewetting line at the corresponding surface. Since the coexisting phases do not possess the symmetry of the Hamiltonian,
$\Delta \mu_{\rm coex}\neq 0$. For short range monomer--wall 
interactions the prewetting--like curves deviate linearly from the bulk coexistence value\cite{HS} and terminate in critical points. 

\vspace*{5mm}
\noindent
    \begin{minipage}[t]{126mm}%
       \mbox{
       ({\bf a})
       \hspace*{-1.5cm}
        \setlength{\epsfxsize}{6.174cm}
        \epsffile{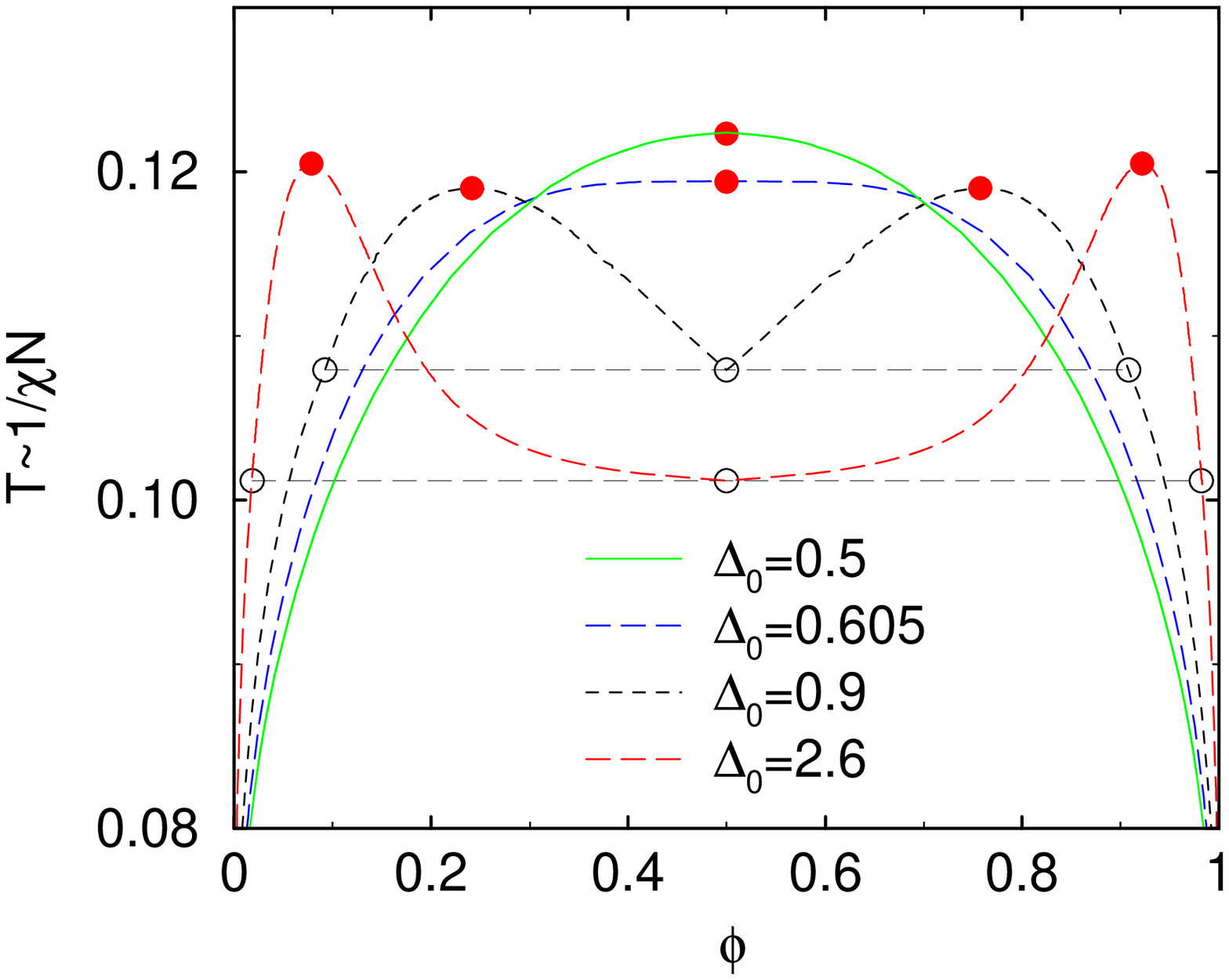}
       ({\bf b})
       \hspace*{-.7cm}
        \setlength{\epsfxsize}{6.3cm}
        \epsffile{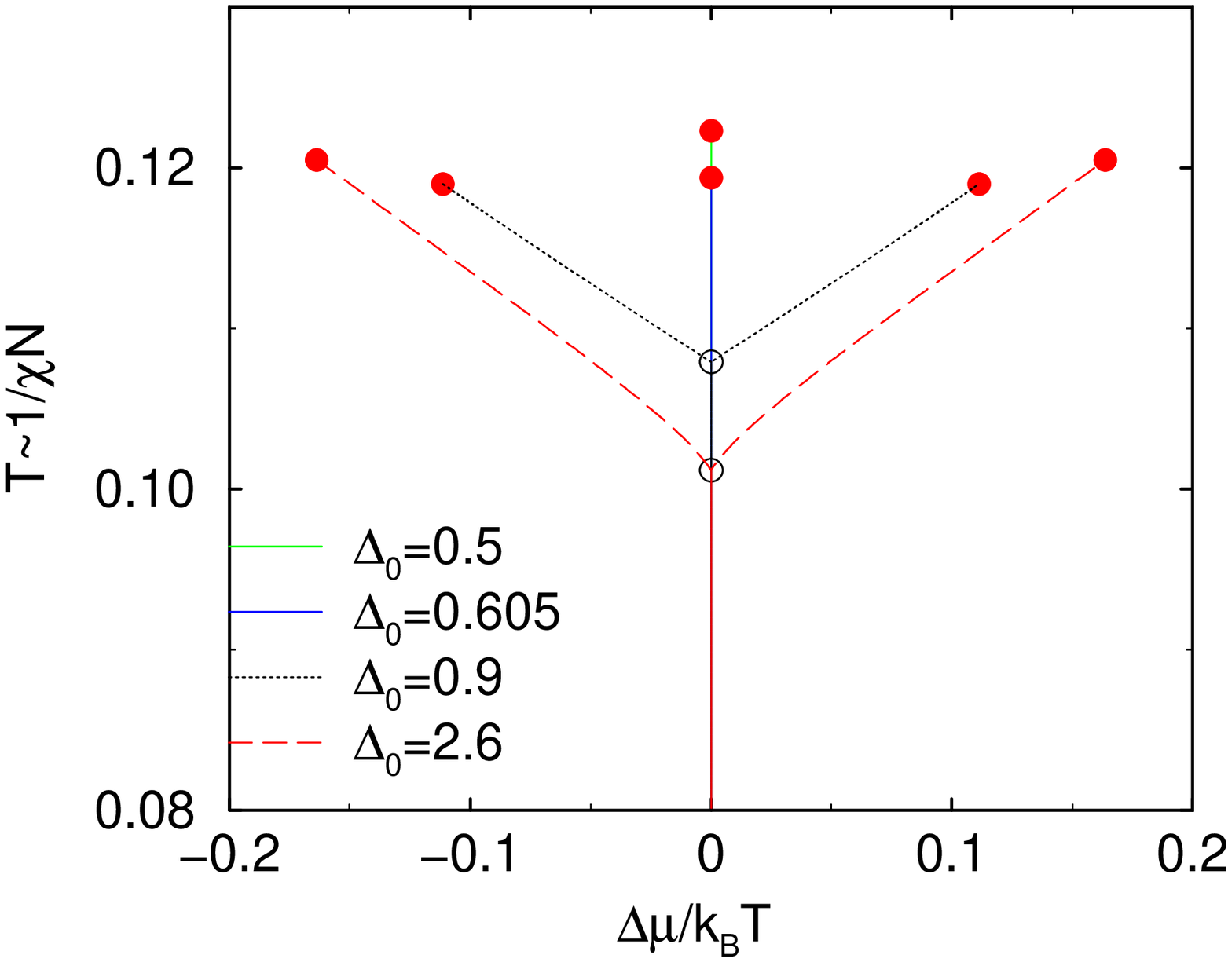}
       }
    \end{minipage}%
    \hfill%
    \begin{minipage}[b]{126mm}%
    {\small
     Fig.2.
     ({\bf a}) Phase diagram for \protect$\Lambda N=0.5$ and various film thicknesses \protect$\Delta_0$. For \protect$\Delta_0=2.6 R_e$ and \protect$0.9R_e$
               the interface localization--delocalization transition is first order, \protect$\Delta_0 = 0.605 R_e$ corresponds to a tricritical
	       transition, while the transition is second order for \protect$\Delta_0 = 0.5R_e$.
     ({\bf b}) Phase diagram as a function of temperature and chemical potential for the same parameters than in ({\bf a}). From \protect{\cite{MSCF2}}
     }
    \end{minipage}%
\vspace*{5mm}

\noindent
The interface localization/delocalization transition can also be changed from first to second order by decreasing the film 
thickness,\cite{SWIFT,MSCF2,MSCF3,MCSCF,B2} because the
effective interactions between the interface and each wall interfere. The three minimum structure of the effective interface potential 
close to the triple point changes to a single minimum as the film thickness becomes comparable to the range of the effective interaction between the 
interface and the wall. The film thickness dependence of the miscibility behavior is presented in Fig.2. For $\Delta_0<0.605 R_e$ the interface 
localization/delocalization transition is second order even though the wetting transition in the semi--infinite system is first order.

Of course, surface interactions in experimental realizations are never strictly antisymmetric or symmetric and it is important which degree of
asymmetry is permissible without loosing the qualitative features of the limiting cases. The phase diagram for non--symmetric boundary fields is 
discussed in Fig.3. 
The right walls attracts the $A$--component of the mixture and the surface fields lead to a first order wetting transition 
in the semi--infinite system. The monomer--wall interactions at the opposite wall are tuned from attracting $A$ (symmetric boundaries, capillary condensation)
to attracting $B$ (antisymmetric boundaries, interface localization/delocal\-ization).\cite{MSCF1}

For symmetric boundaries the critical point is shifted to lower temperatures and higher concentration of the $A$ species attracted by
both surfaces. The coexistence value of the chemical potential $\Delta \mu_{\rm coex}$ is shifted to values disfavoring the $A$ component
and the shift is roughly proportional to the inverse film widths (Kelvin equation). Above the wetting transition temperature an almost pure
$A$--rich phase coexist with a $B$--rich phase. In the latter phase, there are thick enrichment layers of $A$ at the walls and the $B$ component 
prevails at the center of the film. When the temperature is lowered towards the wetting transition temperature $T_{\rm wet}$, the thickness of the 
enrichment layers rapidly decreases. This gives rise to a convex curvature of the $B$--rich binodal slightly above $T_{\rm wet}$.
Below $T_{\rm wet}$ the coexisting phases are almost pure , i.e., the enrichment layers in the $B$--rich phase are negligible and $\Delta \mu_{\rm coex}$
is independent of temperature. 

\vspace*{5mm}
\noindent
    \begin{minipage}[t]{126mm}%
       \mbox{
       ({\bf a})
       \hspace*{-1.5cm}
        \setlength{\epsfxsize}{6.4cm}
        \epsffile{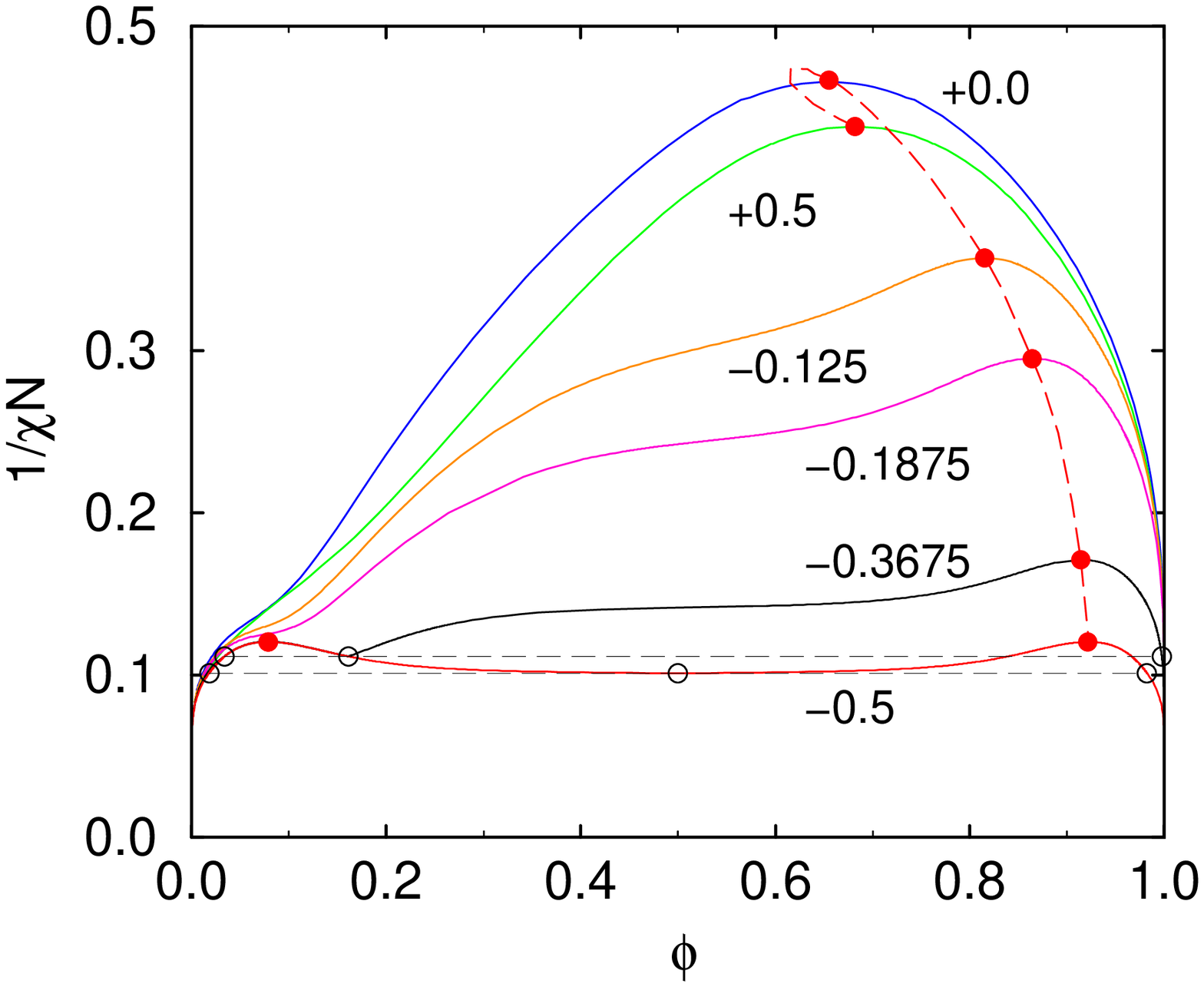}
       ({\bf b})
       \hspace*{-.9cm}
        \setlength{\epsfxsize}{6.4cm}
        \epsffile{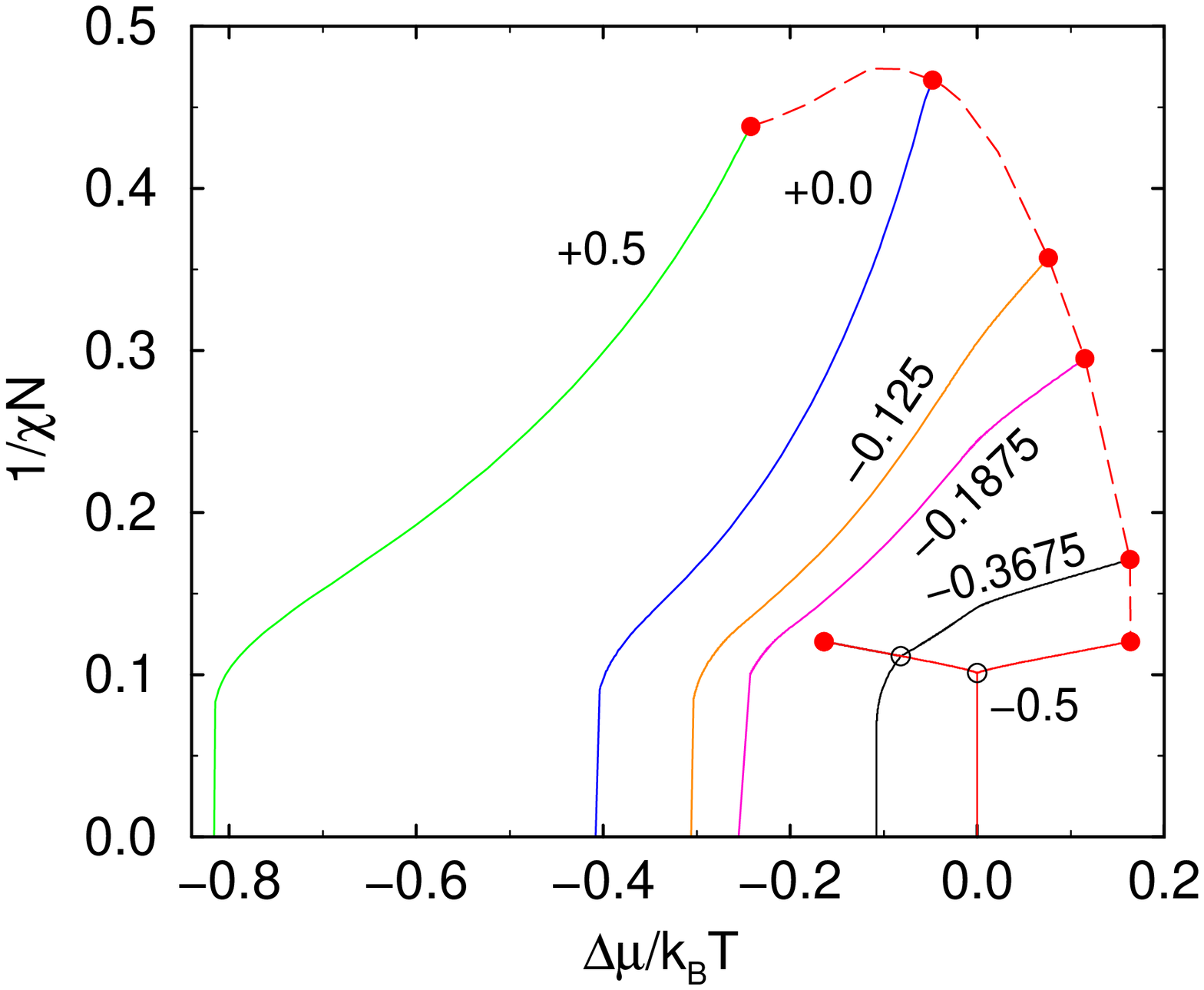}
       }
    \end{minipage}%
    \hfill%
    \begin{minipage}[b]{126mm}%
    {\small
    Fig.3.
    ({\bf a}) Binodals for $\Delta_0=2.6 R_e$ and $\Lambda_1 N=0.5$. $\Lambda_2 N$ varies as
              indicated in the key.  The dashed curve shows the location of the critical points. Filled circles
	      mark critical points, open circles/dashed horizontal lines denote three phase coexistence for
	      $\Lambda_2 N=-0.3675$ and $-0.5$. The inset presents part of the phase boundary for antisymmetric boundaries.
    ({\bf b}) Coexistence curves in the $\chi N$--$\Delta\mu$ plane. $\Lambda_2 N$ varies
              according to the key. The ``quasi--prewetting''
	      lines for $\Delta \mu<0$ and $\Lambda_2N=-0.3675$ and $-0.5$ are indistinguishable, because they are
	      associated with the prewetting behaviour of the surface with interaction $\Lambda_1N=+0.5$. From\protect{\cite{MSCF1}}
    }
    \end{minipage}%
\vspace*{5mm}

\noindent

As we reduce the preference of the left wall for the $A$ component the system becomes more symmetric and the critical point shifts to
more symmetrical composition and higher temperature (i.e., closer to the bulk critical point at $\chi N=2$ and $\phi=1/2$). If we make the left wall repulsive for the 
$A$ component (attractive for the $B$--component)
the character of the transition gradually changes from a bulk--like unmixing transition, where the composition of the two phases varies little spatially
across the film but differs between the two phases, to a prewetting--like transition, where an interface runs parallel to the walls and the distance
between the wall and the interface is the order parameter. As this change occurs the critical temperature (composition) passes through a maximum (minimum).
Upon approaching the strictly antisymmetric limit the coexistence curve (at low temperatures) approaches the bulk value. When it intersects with the
prewetting line of the left wall, which attracts the $B$--species, a second two phase region opens between a thin and a thick enrichment layer at the
left wall.

\section{Monte Carlo simulations}
The self--consistent field calculations presented in the previous section neglect fluctuations. In a cylindrical pore, fluctuations destroy a true phase transition. In a thin film
they change the universality class of the critical points from 3D Ising--like in the bulk to 2D Ising--like in a film. Moreover, there is an strong
interplay between the wetting behavior of the semi--infinite system and the phase diagram of a film with antisymmetric boundaries and interface fluctuations
(i.e., capillary waves) might modify the predictions of the mean field theory.

Being interested in the universal features of the phase diagram we have investigated a coarse grained lattice model of a binary polymer blend\cite{MREV} via
Monte Carlo simulations. In the framework of the bond fluctuation model\cite{BFM} each monomer occupies the corner of a unit cell of a simple cubic lattice from 
further occupation. Monomers along the polymer are connected via bonding vectors of length $2,\sqrt{5},\sqrt{6},3$ or $10$ in units of the lattice spacing.
Monomers interact through a square well potential which comprises the 54 nearest sites on the lattice. A contact between like monomers lowers the energy of
the system by an amount $\epsilon$ (measured in units of $k_BT$), while a contact between different species increases the energy by the same amount.
These interactions set a temperature scale and lead to a liquid--liquid phase separation at $T=1/\epsilon_c=69.3(3)$ in the bulk.\cite{HPD2}
The parameter $\epsilon$ is related to the Flory--Huggins parameter via $\chi = 2 z_c \epsilon$, where $z_c=2.65$ denotes the number of monomers of
other chains in the range of the square well potential.\cite{MREV}

We work at a chain length $N=32$ and monomer number density $\rho=1/16$. The molecules end--to--end distance is $R_e \approx 17$ in units of the lattice spacing.
We study thin film of geometry $L \times L \times \Delta$. Periodic boundary conditions are applied in the two lateral directions, while there are hard impenetrable
walls a distance $\Delta$ apart. Monomers in the two layers nearest to the walls interact with the boundaries. An $A$--monomer close to the right wall decreases the
energy by $\epsilon_w=0.16$, while a $B$--monomer increases the energy by the same amount. For symmetrical walls the interactions at the left wall are identical 
to the right wall;
for antisymmetrical walls $B$--monomers are attracted by the left wall and $A$--monomers repelled. Using the Young equation we have determined the wetting transition
temperature to $T_{\rm wet} = 14.1(7)$.\cite{WET} The wetting transition occurs well inside the strong segregation limit and is of first order.

We use the semi--grandcanonical ensemble, i.e., we fix the temperature and the exchange potential $\Delta \mu$ and monitor the composition $\phi$ and its 
fluctuations. In addition to Monte Carlo moves which update the conformations of the polymers on the lattice (local hopping attempts and slithering--snake--like
motions) we try to change the chain species $A \rightleftharpoons B$ at fixed conformation. This semi--grandcanonical simulation scheme is employed in junction 
with reweighting methods as to encourage the system to sample all compositions with roughly equal probability.

\vspace*{5mm}
\noindent
    \begin{minipage}[t]{63mm}%
       \mbox{
        \setlength{\epsfxsize}{6.7cm}
	\epsffile{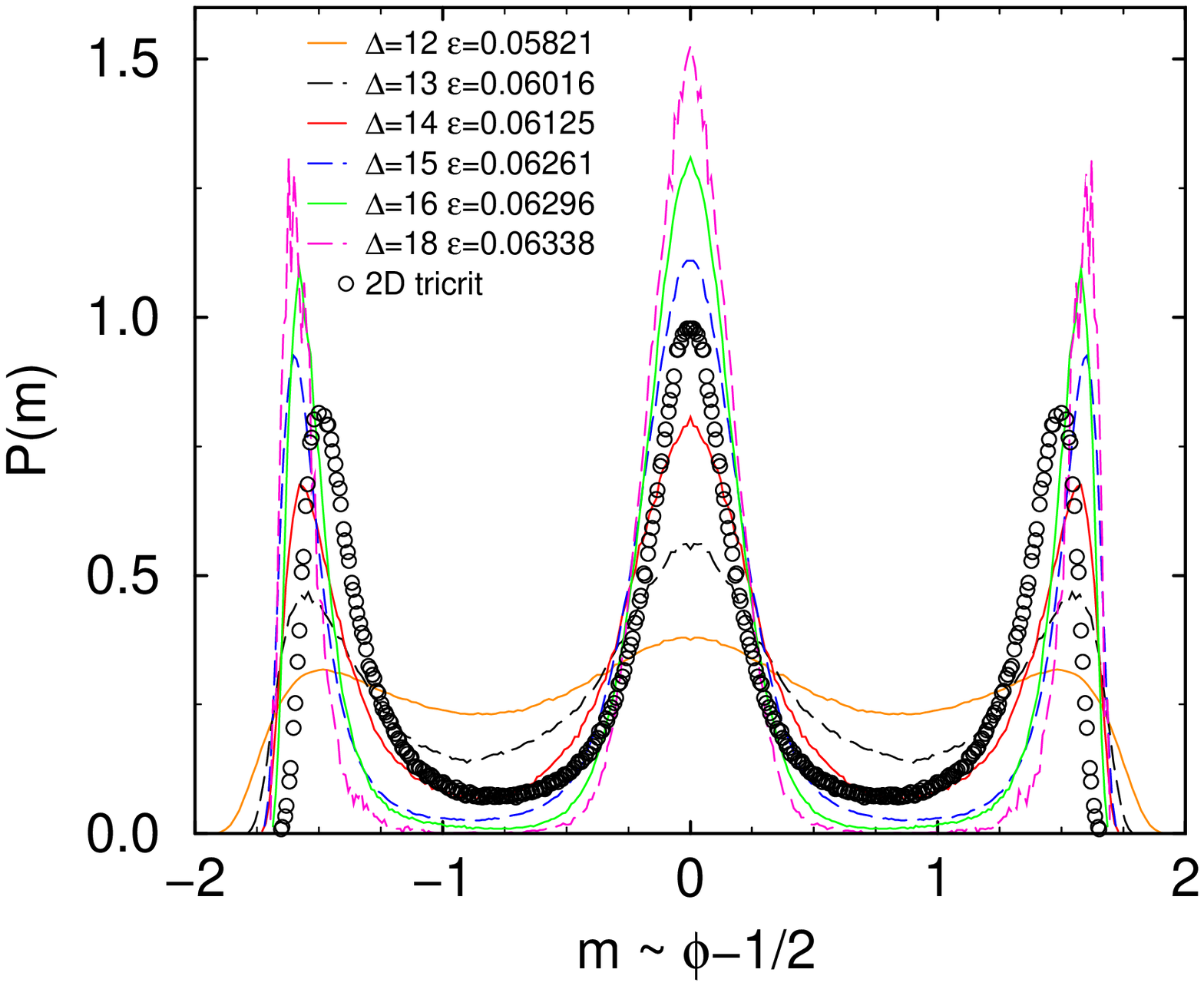}
       }
    \end{minipage}%
    \hspace*{1cm}
    \begin{minipage}[b]{50mm}%
    {\small 
    Fig.4. Probability distribution of the composition scaled to unit norm and variance for various film thickness $\Delta$ ($R_e=17$
    in units of the lattice spacing) and temperatures as indicated in the key. The probability distribution of the 2D tricritical universality 
    class is represented by circles. From \protect{\cite{MCSCF}}
    }
    \end{minipage}%
\vspace*{5mm}

\noindent
In the simulations we monitor the joint distribution of the composition, energy and monomer--wall interaction; and we use finite size scaling techniques 
and histogram reweighting methods to locate the critical point. As an example Fig.4, presents the probability distribution of the composition as a function of 
the film thickness in the vicinity of the tricritical point. The distribution shows a clear three peak structure. The temperature is adjusted such that the
ratio of the heights of the central and outer peaks equals 1.2. This value corresponds to the ratio of the distribution of the tricritical universality
class. The latter quantity has been measured in simulations of the Blume--Capel model at its tricritical point.\cite{NIGEL} To compare the distributions without
adjustable parameter we scale them to unit norm and variance. For film thickness $\Delta=14 \approx 0.82 R_e$ the shape
of the distribution matches closely the universal function and this holds also true for larger systems. For $\Delta< 0.82 R_e$ we find a second order interface 
localization/delocalization transition, while there is a first order transition for $\Delta>0.82 R_e$.

The probability distribution of the composition yields information about the interaction $g(l)$ between the walls and the interface. Since the wetting transition
occurs in the strong segregation limit, the coexisting phases in the bulk are almost pure $\phi_{\rm coex}^{\rm bulk} \approx 0$ or $1$ and ``bulk''--like composition fluctuations
can be neglected. Hence, the distance $l$ between the wall and the interface is given by $l = \Delta \phi$ and the effective interface potential $g(l)$ can
be measured in the Monte Carlo simulations according to $g(l)=- k_BT/L^2 \ln P(\phi)$, where $P(\phi)$ denotes the probability distribution of the composition.

\vspace*{5mm}
\noindent
    \begin{minipage}[t]{65mm}%
       \mbox{
        \setlength{\epsfxsize}{6.7cm}
        \epsffile{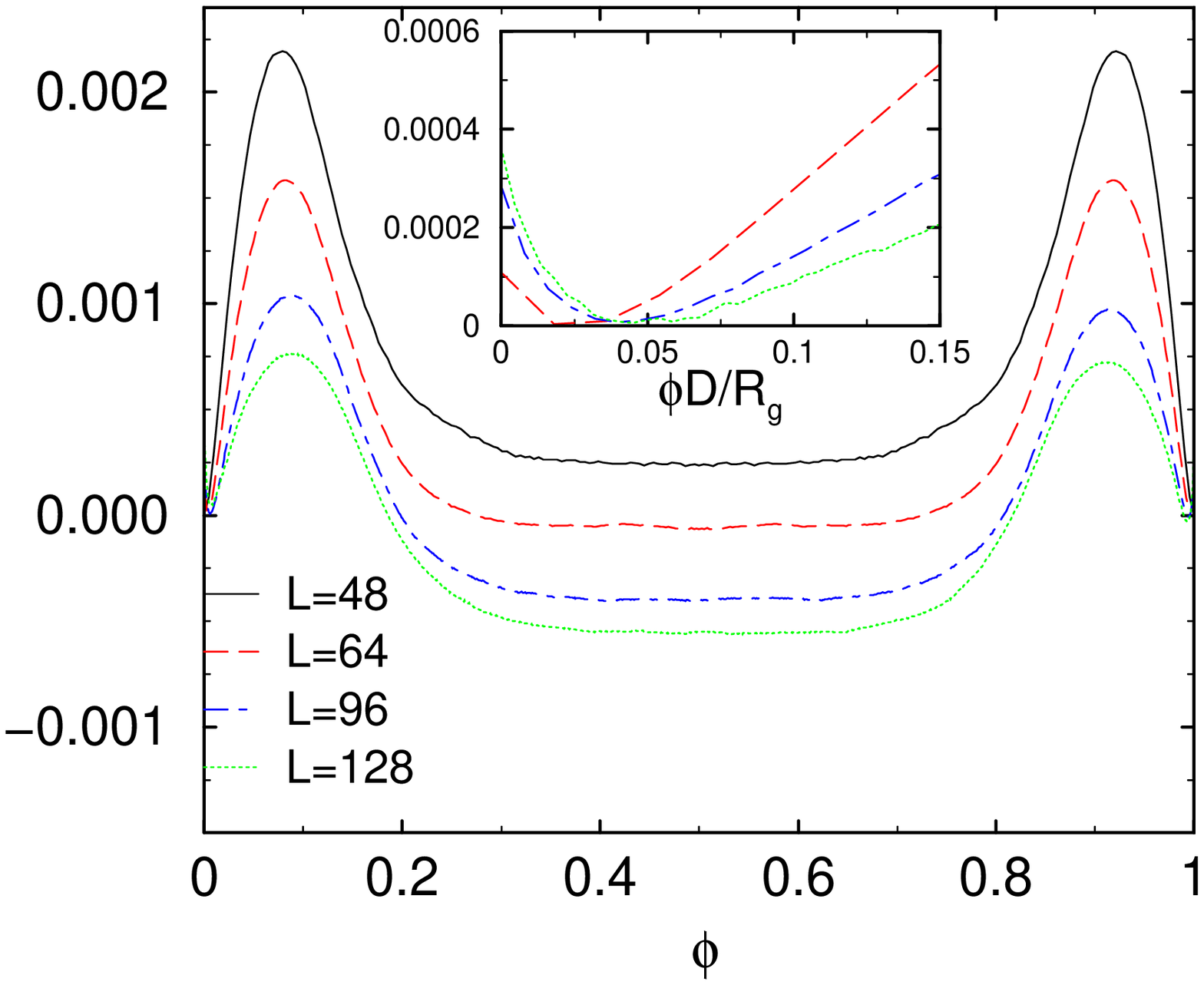}
       }
    \end{minipage}%
    \hfill%
    \begin{minipage}[b]{50mm}%
    {\small
    Fig.5. Dependence of the effective interface potential $g(l)/k_BT$ on the lateral film extension $L$ for $\Delta=48 \approx 2.8 R_e$ 
    and $\epsilon=0.069$. The inset presents an enlarged view on the minimum of the localized state. From \protect\cite{MCSCF}
    }
    \end{minipage}%
\vspace*{5mm}

\noindent
The result for the effective interface potential $g(l)$ in the vicinity of the triple point is presented in Fig.5. The three minima correspond to the
$A$--rich phase, the phase with the delocalized interface, and the $B$--rich phase. Unlike the situation at the tricritical point the position of 
the minima does not depend on the lateral system size. However, the effective interface potential does depend on $L$; the minima broaden upon increasing
the lateral system size $L$ and the free energy of the delocalized state decreases with respect to the localized ones. The dependence of $g(l)$ on the
lateral system size gives rather direct evidence for a renormalization of the effective interface potential by interface fluctuations\cite{LIPOWSKI,BREZIN} in the framework of
a microscopic model. The interface in the simulations is not ideally flat, but there are long wavelength fluctuations of the local interface position (i.e.,
capillary waves). Since the interface is not unconstraint, the interface potential imparts a parallel correlation length onto the capillary waves.
$\xi_\| \sim \sqrt{{\rm d}g^2/{\rm d}^2\phi}$. This parallel correlation length is larger in the delocalized state than in the localized one. For the
parameters of the simulation the lateral system size and the parallel correlation lengths are of the same order of magnitude.
For lateral distances smaller than $\xi_\|$ the local position fluctuates like a free interface, for 
lateral distance that exceed $\xi_\|$ interface fluctuations are strongly suppressed.\cite{MICHAEL} In the simulations the lateral system size $L$ also cuts off
interface fluctuations when $L < \xi_\|$.\cite{ANDREAS} 
Interface fluctuations reduce the free energy of the system and the minima corresponding to the delocalized state benefits more from an increase of 
the lateral system size. This rationalized qualitatively the effect observed in Fig.5 and quantitative description shall be presented elsewhere.\cite{MCSCF}
The effect is important for accurately locating the triple temperature.

\vspace*{5mm}
\noindent
    \begin{minipage}[t]{126mm}%
       \mbox{
       ({\bf a})
       \hspace*{-1.5cm}
        \setlength{\epsfxsize}{6.8cm}
        \epsffile{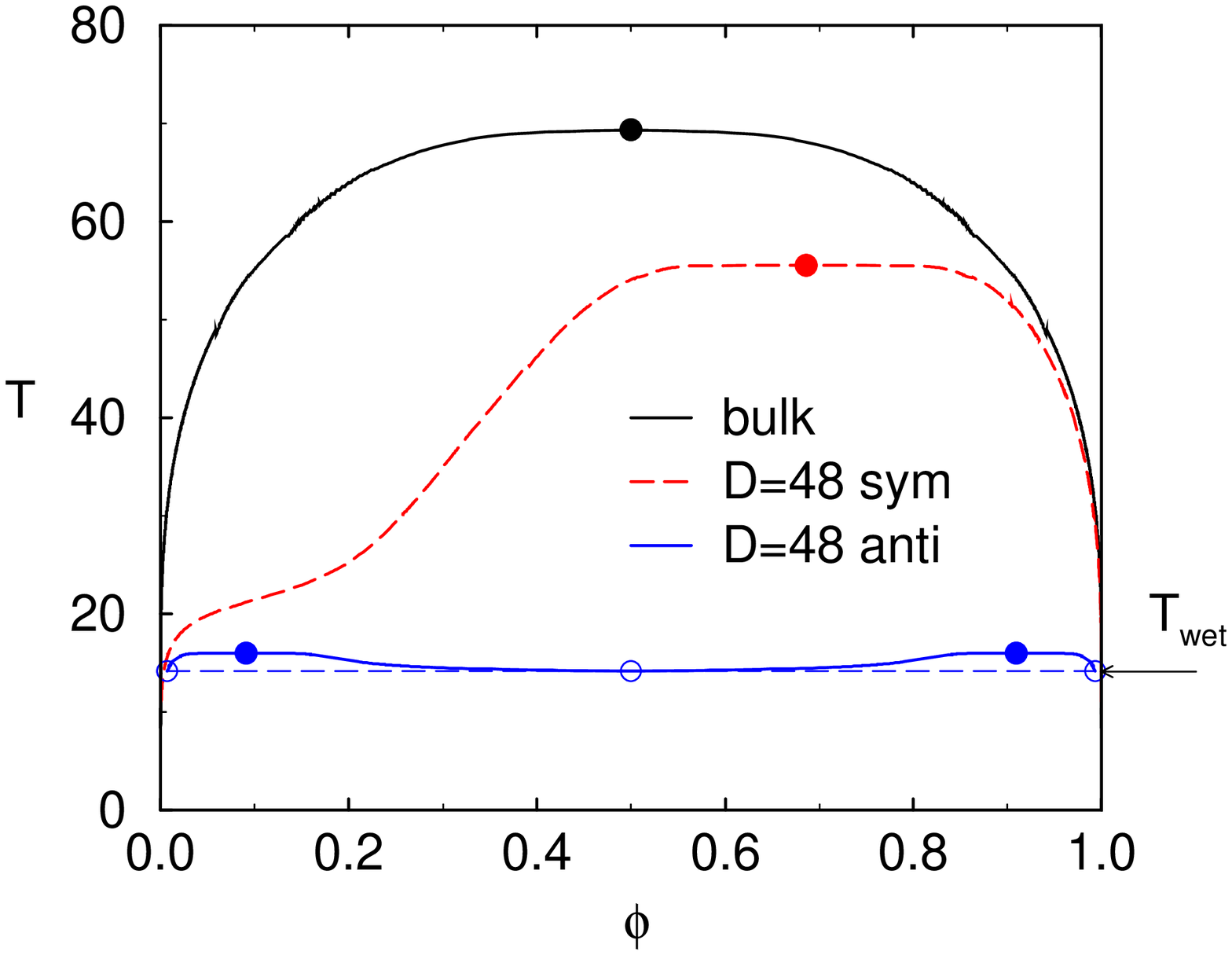}
       ({\bf b})
       \hspace*{-.9cm}
        \setlength{\epsfxsize}{6.4cm}
        \epsffile{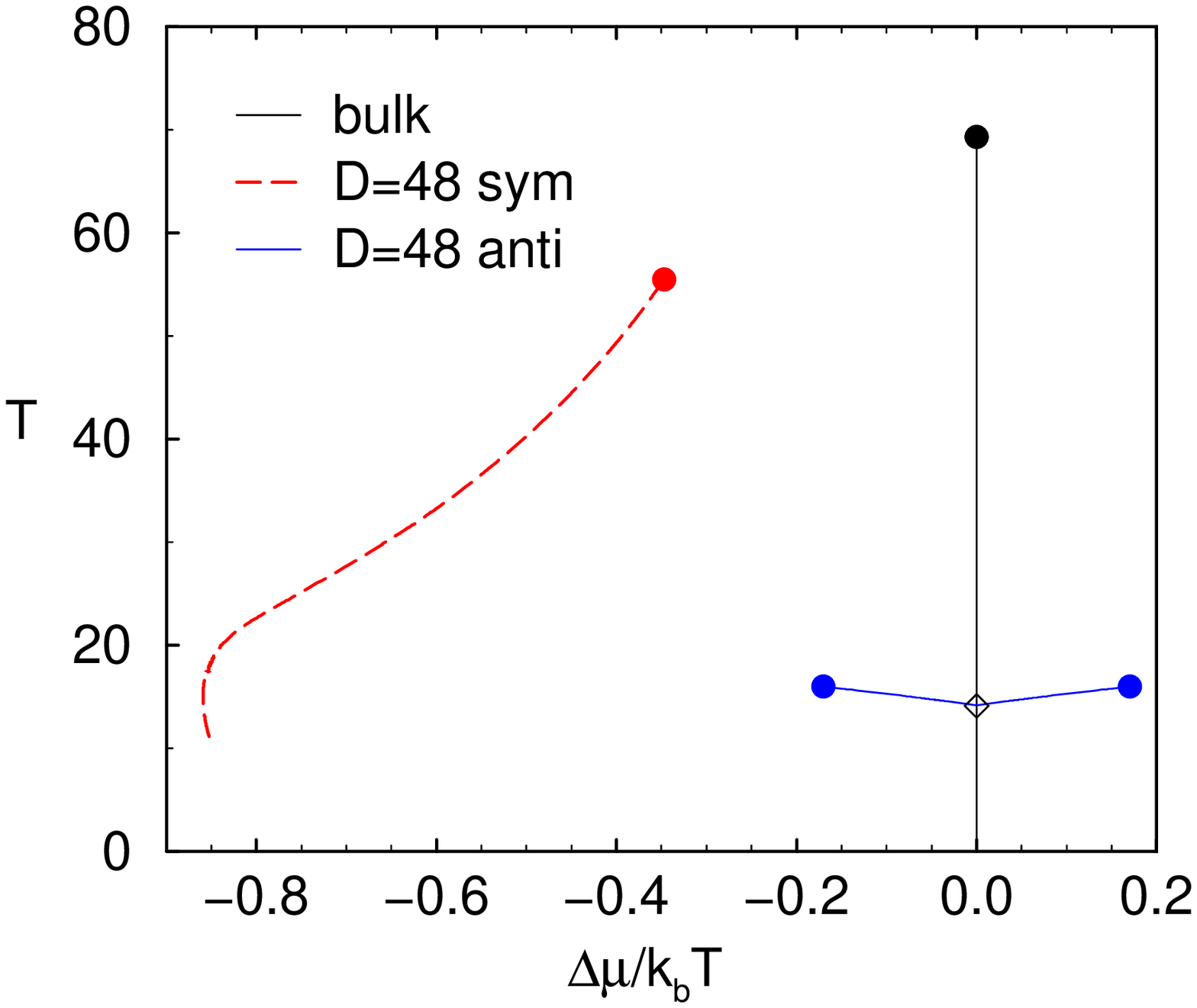}
       }
    \end{minipage}%
    \hfill%
    \begin{minipage}[b]{126mm}%
    {\small
    Fig.6.
    Phase diagram in terms of composition and temperature ({\bf a}) and exchange potential and temperature ({\bf b}) for film thickness
    $\Delta 48 \approx 2.8 R_e$. From \protect\cite{MCSCF}
    }
    \end{minipage}%
\vspace*{5mm}

\noindent
The phase diagrams for $\Delta \approx 2.8 R_e$, as revealed by Monte Carlo simulations in Fig.6, confirm the qualitative predictions of the
self--consistent field calculations. For symmetric boundary conditions there is a small shift of the critical point to lower temperatures and higher 
concentration of the $A$ species prefered by the walls. The ``bulge'' of the $A$--poor binodal is a consequence of the vicinity of the prewetting 
critical point.\cite{WET} If the film thickness were larger the coexistence value $\Delta \mu_{\rm coex}$ would be smaller (Kelvin equation) and the coexistence
curve would intersect the prewetting line. In this case there would be a two phase coexistence region also in the case of symmetric boundaries.\cite{WET,TRIPLE1}
In the antisymmetric case the phase diagram comprises two critical points -- the analogs of the prewetting critical points at each surface.
The concomitant miscibility gaps joint to form a triple point ultimately above the temperature of the first order wetting transition temperature
of the semi--infinite system. Unlike the mean field predictions, however, the shape of the binodals is much flatter reflecting the 2D Ising behavior close to
the critical points.

\section{Discussion}
We have calculated the phase diagram of a symmetric polymer mixture confined into a thin film in mean field approximation
and by Monte Carlo simulations. The mean field calculations reveal a rich interplay between the phase behavior in confined 
geometry and the wetting behavior of the semi--infinite system and the general features of the phase behavior are confirmed
by our Monte Carlo results. However, fluctuations result in two modifications of the mean field results:
(i) In the vicinity of the critical point we observe 2D Ising critical behavior with much flatter binodals than the parabolic 
    binodals of the mean field universality class. 
(ii) The effective interface potential is renormalized by capillary waves. This leads, e.g., to a systematic overestimation of 
    the triple temperature by the mean field treatment.

Qualitatively the interplay between the prewetting behavior and the phase diagram in a film with antisymmetric boundaries is 
not specific to polymer blends but is rather characteristic of all binary mixtures. We expect, however, polymer mixtures, be particluarly
suitable model systems for exploring these effects experimentally.

If we measure the incompatibility of the species/temperature by $\chi N$ and the length scale in units of the end--to--end distance 
$R_e = b \sqrt{N}$, the typical scale of the free energy in a volume $R_e^3$ is given by $\sqrt{\bar N}$, where where the reduced chain length 
$\bar{N}=(\rho R_e^3/N)^2$  measures the degree of interdigitation. 

The importance of fluctuations close to the critical point can be gauged by the Ginzburg criterium. For binary polymer blends one finds 
that fluctuations are important in the range $|1-\chi_c N/\chi N| \ll Gi$ with Ginzburg number $Gi \sim 1/\bar N$. The importance of interface
fluctuations can be described by the capillary parameter\cite{OMEGA1,OMEGA2}
\begin{equation}
\omega=\frac{k_BT \lambda^2}{ 4\pi\sigma_{AB}} = \frac{1}{4 \pi \sqrt{\bar N}} (\lambda R_e)^2 \frac{k_BT \sqrt{\bar N}}{\sigma R_e^2}
\end{equation}
where $1/\lambda$ denotes the characteristic length of the interface potential and $\sigma$ is the interface tension. $\lambda R_e$ and 
$\sigma R_e^2/\sqrt{\bar N} k_BT$ are functions of $\chi N$ only. Upon increasing the degree of interdigitation $\bar N$, the capillary parameter
$\omega \sim 1/\sqrt{\bar N}$ decreases.

In the limit of long chains (or strong interdigitation) these fluctuation effects are suppressed and the self--consistent field theory is
believed to describe the confined blend quantitatively appropriately. For the bond fluctuation model $N=32$ corresponds to
$\bar N \approx 91$ and this chain length of our coarse grained model corresponds roughly to 100-150 repeat units of a real polymer. 
Since fluctuations do not alter the qualitative mean field scenario for the rather short chains investigated by Monte Carlo simulations, 
we expect this a fortiori for experimental systems. Furthermore the wetting
transition in binary polymer mixtures occurs far below the critical point. Bulk--like composition fluctuations are only of minor
importance; the systems are well describable via effective interface Hamiltonians and the effect of interface fluctuations is clearly
observable. Additionally, the size of the enrichment layers is set by $R_e$ and, hence, is much larger than in atomic liquids.
Sophisticated profiling techniques (e.g., nuclear reaction analysis, neutron reflectometry or secondary ion mass spectrometry)
are available to accurately determine the thickness of wetting layers in experiments.

Indeed, experiments on polymer mixtures were the first to investigate fluctuation effects in the delocalized state\cite{KLEIN,SF}
and the wetting transition in polymer blends has been observed experimentally.\cite{EXP2,EXP3} Our results imply, for instance, that 
ultrathin enrichment layers with a thickness $l \approx R_e/5$ be unstable in the temperature range $T_{\rm wet} < T < T_{\rm prewet}$.
Such instabilities slightly above $T_{\rm wet}$ have been observed for compressible one--component polymer films.\cite{EXP1}

Of course, both the self--consistent field calculations and the Monte Carlo simulations deal with highly idealized models. The effect of 
architectural asymmetry between the constituents of the mixture or the role of long--range van der Waals interactions between the monomers 
and the surfaces has not be considered.

\subsection*{Acknowledgment}
It is a great pleasure to thank N.B. Wilding for providing the scaling function of the 2D tricritical universality class (Fig.4).
Financial support was provided by the DFG under grant Bi314/17 in the priority program ``wetting and structure formation at interfaces''
and by the DAAD/PROALAR2000.

\end{document}